\documentclass[aps,prd,twocolumn,superscriptaddress,nofootinbib,10pt]{revtex4-1}
\usepackage[paperwidth=21cm,paperheight=29.7cm,top=2.54cm,bottom=2.54cm,left=2cm,right=2cm]{geometry}
\usepackage[colorlinks,linkcolor=blue,anchorcolor=blue,citecolor=blue,urlcolor=blue]{hyperref}
\usepackage{graphicx,subfigure}
\usepackage[figuresright]{rotating}
\usepackage{amsmath,amsfonts,amssymb,bm}
\usepackage{array,enumitem,multirow}
\usepackage{acronym}
\newcommand{\be}{\begin{equation}}
\newcommand{\ee}{\end{equation}}
\newcommand{\bea}{\begin{eqnarray}}
\newcommand{\eea}{\end{eqnarray}}

\newcommand{\TQC}{MOE Key Laboratory of TianQin Mission, TianQin Research Center for Gravitational Physics \&  School of Physics and Astronomy, Frontiers Science Center for TianQin, Gravitational Wave Research Center of CNSA, Sun Yat-sen University (Zhuhai Campus), Zhuhai 519082, China.}
\newacro{GR}{general relativity}
\newacro{GW}{Gravitational wave}
\newacro{MGT}{modified graivty theory}
\newacro{PN}{post-Newtonion}
\newacro{ppE}{parameterized post-Einsteinian}
\newacro{GCB}{galactic ultracompact binary}
\newacro{SBHB}{stellar-mass black hole binary}
\newacro{MBHB}{massive black hole binary}
\newacro{IMBHB}{intermediate-mass black hole binary}
\newacro{EMRI}{extreme mass ratio inspiral}
\newacro{IMRI}{intermediate mass ratio inspiral}
\newacro{SGWB}{stochastic gravitational wave background}
\newacro{CE}{Cosmic Explorer}
\newacro{ET}{Einstein Telescope}
\newacro{LISA}{Laser Interferometer Space Antenna}
\newacro{EdGB}{Einstein-dilaton Gauss-Bonnet}
\newacro{dCS}{dynamic Chern-Simons}
\newacro{SNR}{signal-to-noise ratio}
\newacro{FIM}{Fisher information matrix}
\newacro{ISCO}{innermost stable circular orbit}
\newacro{LVK}{LIGO, Virgo, and KAGRA Collaboration}
\newacro{TDI}{Time Delay Interferometry}
\newacro{NC}{space-time non-commutativity}
\newacro{PSD}{power spectral density}

\begin{document}

	\title{Testing space-time non-commutativity with TianQin}
	
	\author{Zeyu Huang}
	\affiliation{\TQC}
	\author{Changfu Shi}
	\email{Email: Shichf6@mail.sysu.edu.cn (Corresponding author)}
	\affiliation{\TQC}
	\author{Xiangyu Lyu}
	\affiliation{\TQC}
	\author{Jianwei Mei}
	\affiliation{\TQC}
	
	\date{\today}

\begin{abstract}
The direct detection of gravitational waves offers a powerful tool to explore the nature of gravity and the structure of space-time. This paper focuses on the capabilities of space-based gravitational wave detectors in testing space-time non-commutativity. Our findings indicate that TianQin has the potential to impose constraints on the non-commutative scale at a sub-Planckian level using massive black hole binaries. Additionally, we have developed a pipeline tailored to this specific topic.
\end{abstract}

\maketitle

\section{\label{sec:level1}Introduction}

The direct detection of \acp{GW} provides a new window into the universe \cite{LIGOScientific:2016aoc}, and a new tool of testing the nature of gravity and the structure of space-time\cite{LIGOScientific:2016lio}. Since the first detection of \ac{GW}, the LIGO, Virgo, and KAGRA Collaboration has announced about 90 \ac{GW} events \cite{LIGOScientific:2018mvr,LIGOScientific:2020ibl,LIGOScientific:2021djp}, a series of work focus on testing the validity of \ac{GR} have been carried out \cite{LIGOScientific:2016lio,Abbott:2018lct,LIGOScientific:2019fpa,LIGOScientific:2020tif,LIGOScientific:2021sio,Perkins:2021mhb, Wang:2021jfc,Niu:2021nic, Wang:2021ctl, Kobakhidze:2016cqh,Yunes:2016jcc}, and the results show that all the GW data are consistent with GR so far.

As one possible way to quantum gravity, the idea of non-commutative space-time gained renewed interest\cite{Snyder:1946qz}. This approach treats space-time coordinates as non-commuting entities \cite{PMIHES_1985__62__41_0,Chamseddine:1992yx,Landi:1997sh}. It introduces a new fundamental scale $\theta_{\mu\nu}$ that quantifies the quantum fuzziness of space-time, also known as the non-commutative scale \cite{Kobakhidze:2016cqh}. Experimental constraints have been placed on those parameters from a range of particle physics measurements at low energies and symmetry violation tests, as well as cosmological observations \cite{Joby:2014oee,Akofor:2008gv,Zhang:2004yu,Piscicchia:2022xra}.

Recently, Kobakhidze et al. have proposed a method to constrain the non-commutative scale using \ac{GW} data \cite{Kobakhidze:2016cqh}. They found that the temporal component of the non-commutative tensor affects the phase evolution of \acp{GW}. And by applying this method to GW150914 data, they found that the constraint on \ac{NC} reaches the Planck scale. For a more general case, in which the preferred direction due to non-commutativity has been relaxed, Jenks et al. derived corrections to the evolution of binary systems \cite{Jenks:2020gbt}, and they applied to GWTC-1 event \cite{LIGOScientific:2018mvr} as well as binary pulsar observations of PSR J0737-3039A/B \cite{Kramer:2006nb}.

With technological advancements and the improvement of experimental instrument precision, the third generation of ground-based detectors such as the Cosmic Explorer \cite{Reitze:2019iox} and Einstein Telescope \cite{Punturo:2010zz}, as well as space-based detectors like TianQin \cite{TianQin:2015yph} and LISA \cite{LISA:2017pwj}, are expected to become operational within the next 10 to 20 years. These detectors are anticipated to provide more competitive tests of the nature of gravity and the structure of space-time \cite{Chamberlain:2017fjl,Shi:2022qno}, potentially leading to more stringent constraints on the NC.

In this paper, we investigated the capacity of TianQin on testing NC with inspiral signal from massive black hole binary. TianQin is a planned space-based \ac{GW} detector slated for launch around 2035 \cite{Luo:2015,TianQin:2020hid,Tan:2020xbm,Ye:2020tze}. Its sensitive frequency band ranges from $10^{-4}$ to 1 Hz \cite{Hu:2018yqb,Zhou:2021psj}, making it well-suited for detecting various GW sources. These include galactic ultra-compact binaries \cite{Huang:2020rjf}, massive black hole binaries \cite{Wang:2019ryf,Feng:2019wgq,Chen:2023qga}, intermediate-mass black hole binaries \cite{Liu:2021yoy}, extreme mass ratio inspirals \cite{Fan:2020zhy}, stellar-mass black hole binaries \cite{Liu:2020eko}, stochastic \ac{GW} backgrounds \cite{Liang:2021bde,Liang:2023fdf}, and possibly unexpected sources \cite{TianQin:2020hid,Fan:2022wio,Wu:2023rpn}. These signals offer rich opportunities for astronomical \cite{Torres-Orjuela:2023hfd}, cosmological \cite{Zhu:2021aat,Zhu:2021bpp,Huang:2023prq,Lin:2023ccz}, and fundamental physics research \cite{Shi:2019hqa,Bao:2019kgt,Zi:2021pdp,Sun:2022pvh,Sun:2024nut,Xie:2022wkx,Shi:2022qno,Wang:2021dwl,Wang:2020jrd,Wei:2022poh}. Our work contributes to this fundamental physics research and also involves corresponding data processing techniques.

Our result indicate that TianQin has the potential to impose constraints on the temporal component of the space-time non-commutative tensor at sub-Planckian scale. To investigate data processing with real data, we conducted Bayesian inference on mock TianQin data involving massive black hole binary, taking into account the orbital motion of the TianQin satellite and the \ac{TDI} response of the detector.

The paper is organized as follows. In Sec. \ref{sec:data}, we introduce the waveform model, the basic of \ac{TDI}, the noise of TianQin, and data analysis methods. In Sec. \ref{sec:result}, we present our main results. And a summary are presented in section \ref{sec:summary}. In this study, we refer to the natural units in which $G_N=\bar{h}=c=1\,$ unless otherwise specified.

\section{Methodology}\label{sec:data}
In this section, we present the main tool utilized in this paper, which includes the waveform of \ac{GR}, the correction introduced by NC, the \ac{TDI} response, the noise characteristics of TianQin, and the data analysis methods: Fisher information matrix and Bayesian inference.

\subsection{Waveform}
Currently, the \ac{GW} waveform of any modified theories of gravity are available in the inspiral stage.
As two components of the binary are widely separated and their velocities are relatively small compared to the speed of light, one can utilize the post-Newtonian approximation to describe the waveform for those binaries with comparable masses. According to \cite{Yunes:2009ke}, the post-Newtonian waveform for inspiral stage in frequency domain can be written as:
\begin{align}
	\label{eq:pn}
	h(f)&=A(f)e^{i\Psi(f)}\notag\\
	A(f)&=(1+\alpha u^a)A(f)_{GR},\notag\\
	\Psi(f)&=\Psi_{GR}(f)+\beta u^b.
\end{align}
Where $\alpha u^a$ and $\beta u^b$ represent the leading order amplitude and phase corrections. The correction order parameters $a=2\rm PN$ and $b=2\rm PN-5$ indicate the post-Newtonian order at which the leading order correction occurs. $\alpha$ and $\beta$ are correction parameters that measure deviations from \ac{GR}, for the case of \ac{GR}: $\alpha=0$, $\beta=0$.
The typical velocity of the system is given by $u=(\pi M \eta^{3/5}f)^{1/3}$, where $M=m_1+m_2$ is the total mass and $\eta=m_1m_2/M^2$ is the symmetric mass ratio of the binary. $m_1$ and $m_2$ are the masses of the major and minor components of the binary, respectively. 
The amplitude and phase of the \ac{GW} waveform predicted by GR, and they are denoted by $A_{GR}(f)$
and $\Psi_{GR}(f)$, respectively. Given that the measurement accuracy of laser interferometric GW detectors for phase is significantly better than that for amplitude, only the phase correction contribution is considered in this paper.

In the simplest realization of NC, the coordinate operators $\hat{x}$ satisfy the canonical commutation relations given by  
\begin{align}
	\label{eq:nc}
	[\hat{x}_\mu, \hat{x}_\nu]=i\theta_{\mu\nu},
\end{align}
where $\theta_{\mu\nu}$ is a real constant anti-symmetric tensor. Similar to the Planck constant, $\theta_{\mu\nu}$ is considered to be a new fundamental physical quantity that quantifies the quantum fuzziness of space-time.

Following  Kobakhidze et al. \cite{Kobakhidze:2016cqh}, the corrections introduced by NC to the  energy-momentum tensor of a binary system can be written as:
\begin{align}
	\label{eq:emt}
	T_{NC}^{\mu\nu}&=\frac{m_1^3\Lambda^2}{8}v_1^\mu v_1^\nu\theta^k\theta^l\partial_k\partial_l\delta^3(x-y_1(t))\notag\\&+1\leftrightarrow 2+T_{GR}^{\mu\nu}.
\end{align}
This correction will lead to modification in phase at 2PN order, and the phase correction can be written as:
\begin{align}
	\label{eq:phase}
	&\Delta\phi_{NC}^{2\rm PN}=-\frac{75}{256}\eta^{-4/5}(2\eta-1)\Lambda^2(\pi M\eta^{3/5}f)^{-1/3},\notag\\
	&\Lambda^2\equiv|\theta^{0i}|^2/(l_pt_p)^2,
\end{align}
where $l_p=\sqrt{\bar{h}G/c^3}\approx1.6\times10^{-35}\rm m$ and $t_p=\sqrt{\bar{h}G/c^5}\approx5.4\times10^{-44}\rm s$ are the Planck length and Planck time, respectively. The parameter $\sqrt{\Lambda}$
represents the ratio of the non-commutative scale to the Planck scale. From Eq.\eqref{eq:pe}, one can conclude:
\begin{align}
	\label{eq:ppenc}
	\beta_{NC}==-\frac{75}{256}\eta^{-4/5}(2\eta-1)\Lambda^2,b_{NC}=-1
\end{align}
where $b_{NC}=-1$ indicates that the leading order correction of the waveform due to NC occurs at the 2PN order.
Using this waveform, they found a constraint $\sqrt{\Lambda}<3.4$ by comparing it with GW150914 data.

As Eq. \eqref{eq:pn} is valid for cases where only the 22-mode are considered, while the contribution of higher-order modes are important for systems with large mass ratios and high eccentricities. Therefore, a natural extension of Eq. \eqref{eq:pn} can be expressed as:
\begin{align}
	\hat{h}_{NC}(f)&=\sum_{lm}^{}\hat{h}_{NC,lm}(f),\\
	\hat{h}_{NC,lm}(f)&=\hat{h}^{GR}_{lm}(f)\exp^{i\beta_{NC,lm}u^{b_{NC,lm}}},
\end{align}
where $\beta_{NC,lm}$ and $b_{NC,lm}$ represent the phase correction parameters and phase correction order parameter for any $lm-$mode, respectively. As demonstrated in \cite{Mezzasoma:2022pjb}, for non-precessing binaries:
\begin{align}
	\label{eq:waveform}
	\beta_{NC,lm}=({\frac{2}{m}})^{b_{NC}/3-1}\beta_{NC},
	b_{NC,lm} = b_{NC},
\end{align}
where $l=2,3,4,\dots$ and $m=0, \pm 1, \pm 2, \dots $ are the harmonic indices.
In this paper, we employ IMRPhenomXHM to generate the \ac{GW} waveform in \ac{GR}. The higher harmonics modes considered include  $(l,|m|)=\{(2,2),(2,1),(3,3),(3,2),(4,4)\}$.

\subsection{Detector response}
In this paper, we use \ac{TDI} to model detector response under the scenario of strictly equal arm lengths for the TianQin satellite. The key steps of this method are outlined in this subsection, and a comprehensive derivation can be found in \cite{Marsat:2020rtl,Marsat:2018oam}.

We start with considering  a laser  frequency shift between the laser links of spacecraft $r$ and $s$:
\begin{align}
	\label{eq:sltd}
	   y_{slr}&=(\mu_{r}-\mu_{s})/\mu\notag\\&=\frac{1}{2} \frac{n_l \otimes n_l}{1-k \cdot n_l} : [h(t-L-k\cdot p_s)
	   -h(t-k\cdot p_r)],
\end{align}
where $k$ denotes the wave propagation vector, $L$ represents the delay along one arm, $n_l$ stands for the link unit vectors, while $p_s$, $p_r$ are the positions of the spacecrafts. 
By incorporating higher harmonics of \acp{GW} and applying a Fourier transformation, the frequency domain representation $\hat{y}_{slr}$ can be formulated as:
\begin{align}
	\label{eq:fsltd}
	\hat{y}_{slr}=\sum_{l,m}\mathcal{T}^{lm}_{slr}(f)\hat{h}_{lm}(f),
\end{align} 
where $\mathcal{T}^{lm}_{slr}$ represents the transform function
\begin{align}
	\label{eq:tf}
	\mathcal{T}^{lm}_{slr}(f)&=\frac{i\pi fL}{2} \text{sinc}[\pi fL(1-k\cdot n_l)] \\
	&\cdot \text{exp}[i\pi f(L+k\cdot(p_r+p_s))]n_l \cdot P_{lm} \cdot n_l,\notag
\end{align}
with $P_{lm}$ represents the polarization tensor associated with the $(l,m)$ mode. This simplified transform function $\mathcal{T}^{lm}_{slr}(f)$
depends on the frequency $f$ and a time-frequency relation $t_f^{lm}$
given by $t^{lm}_f=-\frac{1}{2\pi} \frac{d\Psi_{lm}}{df}$. 
It is worth noting that the detector response varies for each harmonic and is closely related to the spin-weighted spherical harmonics.

For a single-arm laser link, the observable $y_{slr}$ is significantly impacted by laser noise. One can mitigate this noise by reconstructing a set of new observables, namely the A,E,T channels, utilizing $y_{slr}$. The TDI channels in the frequency domain can be expressed as:
\begin{align}
	\label{eq:aetf}
	\hat{a} &= (1+z)(\hat{y}_{31}+\hat{y}_{13})-\hat{y}_{23}-z\hat{y}_{32}-\hat{y}_{21}-z\hat{y}_{12}, \notag \\
	\hat{e} &= \frac{1}{\sqrt{3}}[(1-z)(\hat{y}_{13}-\hat{y}_{31})+(2+z)(\hat{y}_{12}-\hat{y}_{32}) \notag\\
	&+ (1+2z)(\hat{y}_{21}-\hat{y}_{23})], \\
	\hat{t} &= \frac{\sqrt{2}}{\sqrt{3}}[\hat{y}_{21}-\hat{y}_{12}+\hat{y}_{32}-\hat{y}_{23}+\hat{y}_{13}-\hat{y}_{31}], \notag
\end{align}
and
\begin{align}
	\label{eq:rescaled}
	\hat{a},\hat{e}&=\frac{\exp^{-2i\pi fL}}{i\sqrt{2}\sin(2\pi fL)}\times \hat{A},\hat{E},\\
	\hat{t}&=\frac{\exp^{-3i\pi fL}}{2\sqrt{2}\sin(\pi fL)\sin(2\pi fL)}\times\hat{T}.
\end{align}
where $z\equiv \exp[2i\pi fL]$ and the rescaling relations given in Eq. \eqref{eq:rescaled} have been applied to eliminate frequency-dependent prefactors that are common to both the signal and noise. These prefactors are oscillatory and have zero-crossings at high frequencies. Now, with these transfer functions, we can obtain the signal after TDI response:
\begin{align}
	\label{eq:tdir}
	\hat{a}, \hat{e}, \hat{t} &= \sum_{l,m}^{}\mathcal{T}^{lm}_{a,e,t}\hat{h}_{lm}.
\end{align}

\subsection{Nosie}
When it comes to noises in the A, E, and T channels, they can be modeled as follows:
\begin{align}
	\label{eq:noise}
	S^{A}{n}, S^{E}{n} &= 8\sin^{2}(2\pi fL) \times S^{a}{n}, S^{e}{n}, \notag\\
	S^{T}{n} &= 32\sin^{2}(\pi fL)\sin^{2}(2\pi fL)S^{t}{n},
\end{align}
where $S^{A}{n}, S^{E}{n}$, and $S^{T}{n}$ are given by:
\begin{align}
	S^{a}{n},S^{e}{n} =& 2(3+2\cos(2\pi fL)+\cos(4\pi fL))S_{\text{acc}}(f) \notag\\
	&+ (2+\cos(2\pi fL))S_{\text{pos}}(f), \notag\\
	S^{t}{n} =& 4\sin^{2}(2\pi fL)S{\text{acc}}(f) + S_{\text{pos}}(f).
\end{align}
According to the current design of TianQin, 
the \ac{PSD} of acceleration noise on the test mass are set to be $S_{\rm acc}=1\times10^{-15}\rm m^2s^{-4} Hz^{-1}$, while the \ac{PSD} of position noise in a single link are set to be $S_{\rm pos}=1\times10^{-24} \rm m^2Hz^{-1} $.

\subsection{Data analysis method}\label{sec:method}
In this paper, two different data analysis methods are used, i.e. Fisher information matrix and Bayesian inference.

For the case of large \ac{SNR} signals and Gaussian noise, the parameter estimation accuracy for any parameter $\theta_\alpha$ of waveform model can be approximated by \cite{Finn:1992,Cutler:1994}
\begin{align}
	\label{eq:pe}
	\delta\theta_\alpha=\sqrt{\langle\delta\theta_\alpha\delta\theta_\alpha\rangle}\simeq\sqrt{(\Gamma^{-1})_{\alpha\alpha}},
\end{align}
where $\langle\dots\rangle$ stands for statistical average, and $\Gamma^{-1}$ is the covariance matrix, which is the inverse of the Fisher information matrix. With the consideration of \ac{TDI} response, the matrix can be calculated by summing of the contribution from A, E, and T channels:
\begin{align}
	\label{eq:fimall}
	\Gamma_{\alpha\beta}=\sum_{N\in{A,E,T}}\Gamma_{\alpha\beta}^{N}\equiv\sum_{N\in{A,E,T}}\big(\frac{\partial h^{N}}{\partial\theta^\alpha}|\frac{\partial h^{N}}{\partial\theta^\beta}\big).
\end{align}
Where $\big(\dots|\dots\big)\big)$ is the inner product, defined as:
\begin{align}
	\label{eq:inner-product}
	(p|q)\equiv 2\int_{f_{low}}^{f_{high}}\frac{p^*q+q^*p}{S_{n}^{N}(f)} df.
\end{align}  
where $p$ and $q$ are any frequency domain signals, and $S_n^{N}(f)$ in are the noises in this channel. And the optimal  \ac{SNR} is defined as the inner product of signal $\rho^2[h]\equiv {(h|h)}$.

In this paper, we deal with a 12-dimensional parameter space.
\begin{align}
	\label{eq:ps}
	\theta=\{M,\eta,D_L,\iota,\chi_1,\chi_2,\lambda,\beta,\psi,t_c,\phi_c,\Lambda\}.
\end{align}
The physical meaning of those parameter are listed in the first and the second column of table. \ref{tab:paras}. Note $\chi_1$ and $\chi_2$ are dimensionless and the spins are assumed to be aligned with the orbital angular momentum. 
\begin{table*}
	\centering
	\caption{The parameters involved in this paper.}
	\renewcommand{\arraystretch}{1.2}
	\begin{tabular}{|c|c|c|c|c|c|}
		\hline
		Symbol & Physical meaning &Source I & Priors for Source I  &Source II & Priors for Source II \\
		\hline
		SNR & Signal to noise ratio & 580 & - & 329 & -\\ \hline
		$M$              & Total mass of the binary & $10^6 \rm M_\odot$ &[998000,1002000]$ \rm M_\odot $& $10^6 \rm M_\odot$&[998000, 1002000]$\rm M_\odot$ \\ \hline
		$\eta$             & Symmetric mass ratio of the binary & 0.24793 & [0.23,0.24999] & 0.08264 & [0.07,0.10]\\ \hline
		$\chi_1$             & Spin of the major component & 0.6 & [-1,1] & 0.6& [-1,1]\\ \hline
		$\chi_2$             & Spin of the minor component & 0.4 & [-1,1] & 0.4 & [-1,1]\\ \hline
		$\lambda$        & Right ascension of source location & 1.2 & [0, $\pi/2$]& 1.2 & [0, $\pi/2$] \\ \hline
		$\beta$          & Declination of the source location & 0.7 &[0, $\pi/2$] & 0.7 & [0, $\pi/2$] \\ \hline
		$\psi$            & Polarization angle & 0.7 &[0, $\pi/2$] & 0.7 & [0, $\pi/2$] \\ \hline
		$\iota$           & Inclination angle & 1.1 & [0, $\pi/2$] & 1.1 &[0, $\pi/2$] \\ \hline
		$\phi_{\rm c}$  & Phase at the reference frequency & 1.2 & [0, $\pi/2$] & 1.2 & [0, $\pi/2$] \\ \hline
		$t_c$             & the reference time & 0 &[-100,100] & 0 & [-100,100] \\ \hline
		$D_L$             & Luminosity distance & 6 Gpc & [5 Gpc,7 Gpc] & 6 Gpc&  [5 Gpc,7 Gpc]\\ \hline
		$\Lambda$ & Parameter from \ac{NC} & 0 & [0,10] & 0 & [0,10]\\ \hline
	\end{tabular}
	\label{tab:paras}
\end{table*}

The actual processing of real \ac{GW} data often relies on Bayesian inference. 
According to the Bayes' theorem,  given the data $d$ and hypothesis $H$, the posterior probability distribution of a set of parameters $\theta$ is given by
\begin{align} p(\theta|d, H)=\frac{p(d|\theta, H) \pi(\theta|H)}{\mathcal{Z}(d|H)}\,,\label{eq:Bayes_theorem}
\end{align}
where $p(d|\theta, H)$ and $\pi(\theta|H)$ are the likelihood and prior. $\mathcal{Z}(d|H)$ is an overall normalization, also called evidence.

With the assumption of stationary and Gaussian noise, the likelihood 
$\mathcal{L}(d|\theta,H)$ in N channel can be written as:
\begin{align}
	\ln \mathcal{L}^N=-\frac{1}{2}(h^N(\theta)-d^N|h^N(\theta)-d^N),
\end{align}
where $d^N=h^N(\theta_{0})+n^N$ is the  data, $h^N(\theta_{0})$ is the signal with injected value $\theta_{0}$ and response with N channel, $n^N$ stands for noise in N channels of TianQin.

By summing the contributions from A, E, and T channels, the log likelihood of model can be written as:
\begin{align}
	\ln \mathcal{L}^{Total}=&\sum_{N\in{A,E,T}}\ln \mathcal{L}^{N}.
\end{align}

\section{result}\label{sec:result}
In this section, we outline the main results, including the testing capacity obtained by Fisher information matrix and the Bayesian inference of mock data.
\subsection{Testing capability}
According to Eq.\eqref{eq:pe}, the accuracy of $\delta\sqrt{\Lambda}$
is primarily influenced by the $\Lambda\Lambda$ component of the Fisher information matrix. This suggests that parameters such as the total mass, symmetric mass ratio, luminosity distance, sky location, and inclination should be carefully considered. Notably, $D_L$
has a linear impact on signal strength, and its influence on $\sqrt{\Lambda}$ estimation accuracy follows a $D_L^{-1/4}$ trend. Although sky location and inclination contribute to the detector response, previous studies \cite{Wang:2019ryf,Huang:2020rjf} have demonstrated that these factors do not significantly alter the outcomes.
Therefore, without loss of generality, we set $D_L=15 \rm Gpc$, $t_c=0$, $\phi_c=0$, $\chi_1=0.7$, $\chi_2=0.5$, $\iota=\pi/3$, $\lambda=1.2$, $\beta=0.7$, $\psi=0.7$.

This paper aims to assess capability of TianQin in testing NC using massive black hole binary. As suggested in \cite{Shi:2022qno}, these systems can place stringent constraints on 2PN corrections. We consider black hole binary systems with masses ranging from $[10^4\rm M_\odot, 10^7\rm M_\odot]$. Additionally, to maintain the validity of the IMRPhenomXHM and waveform correction calculations, this study limits the mass ratio ($q=m_1/m_2$, and $m_1>m_2$) to a maximum of 10.

\begin{figure}[t]
	\centering
	\includegraphics[width=0.54\textwidth]{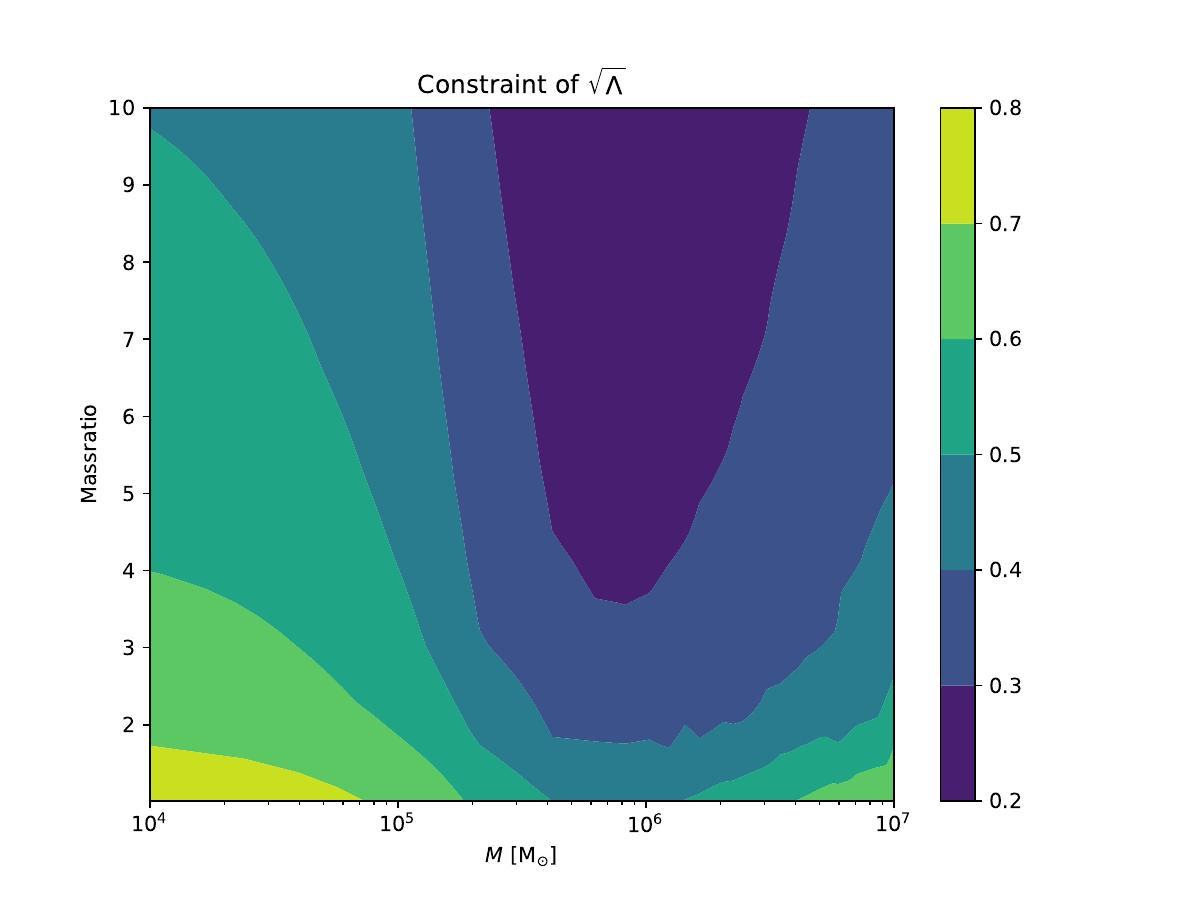}
	\caption{The dependences of $\delta\sqrt{\Lambda}$ on $M$ and $q$.}\label{fig:NC-constrain}
\end{figure}

The detailed result of $\delta\sqrt{\Lambda}$ is displayed in Fig.~\ref{fig:NC-constrain}. Despite some optimism regarding the selection of the source location, considering the impact of source parameters on the results, one can also infer from Fig. \ref{fig:NC-constrain} that TianQin has the potential to constrain NC to $\sqrt{\Lambda}<0.3$. This indicates that the temporal part of the non-commutative tensor $\theta_{\mu\nu}$ can be constrained more tightly than the Planck scale by a order of magnitude.
Furthermore, Fig. \ref{fig:NC-constrain} also reveals that the most stringent constraints are achievable for total masses around $10^6 \rm M_\odot$ with asymmetric mass ratios.

\subsection{Bayesian inference}
To address the future need for real data processing, we have developed a pipeline aimed at exploring NC using space-based \ac{GW} detector. To assess its performance, we have conducted an Bayesian analysis on simulation data from massive black hole binary.

Using the waveform predicted by GR and the noise properties of the TianQin detector, we simulate the data of massive black hole binary detected by TianQin. We employ the IMRPhenomXHM waveform to represent the GW signal, and the noise is generated based on the \ac{PSD} of noise in each channel (eq. \ref{eq:noise}). We consider two typical sources with mass ratios of q=1.2 (similar to GW150914) and q=10. The detailed source parameters for these two cases are listed in the third and fifth columns of Table \ref{tab:paras}.

For Bayesian inference, we employ the Bilby and Dynesty frameworks. We set the number of live points to 1500 and the logarithmic spacing between samples to 0.01. We test two waveform models for our Bayesian analysis: Eq. \eqref{eq:waveform} and IMRPhenomXHM. For all parameters except $D_L$, $\Lambda$ and $\beta$, which are assumed to be spatially uniform, we employ a uniform prior distribution. The specific ranges of the prior distributions for all 12 parameters are provided in Table \ref{tab:paras}, columns 4 and 6. These ranges are designed to encompass more than 10 times the standard deviations derived using the Fisher information matrix.
It is worth noting that we reparameterize the spin parameters  $\chi_a,\chi_l=(\chi_1 +\chi_2)/2,(\chi_1 -\chi_2)/2$ instead of using  $\chi_1$ and $\chi_2$ directly to improve sample efficiency and reduce computational time, as suggested by \cite{Lyu:2023ctt} and \cite{Marsat:2020rtl}. 

\begin{figure*}[t]
	\centering
	\includegraphics[width=\textwidth]{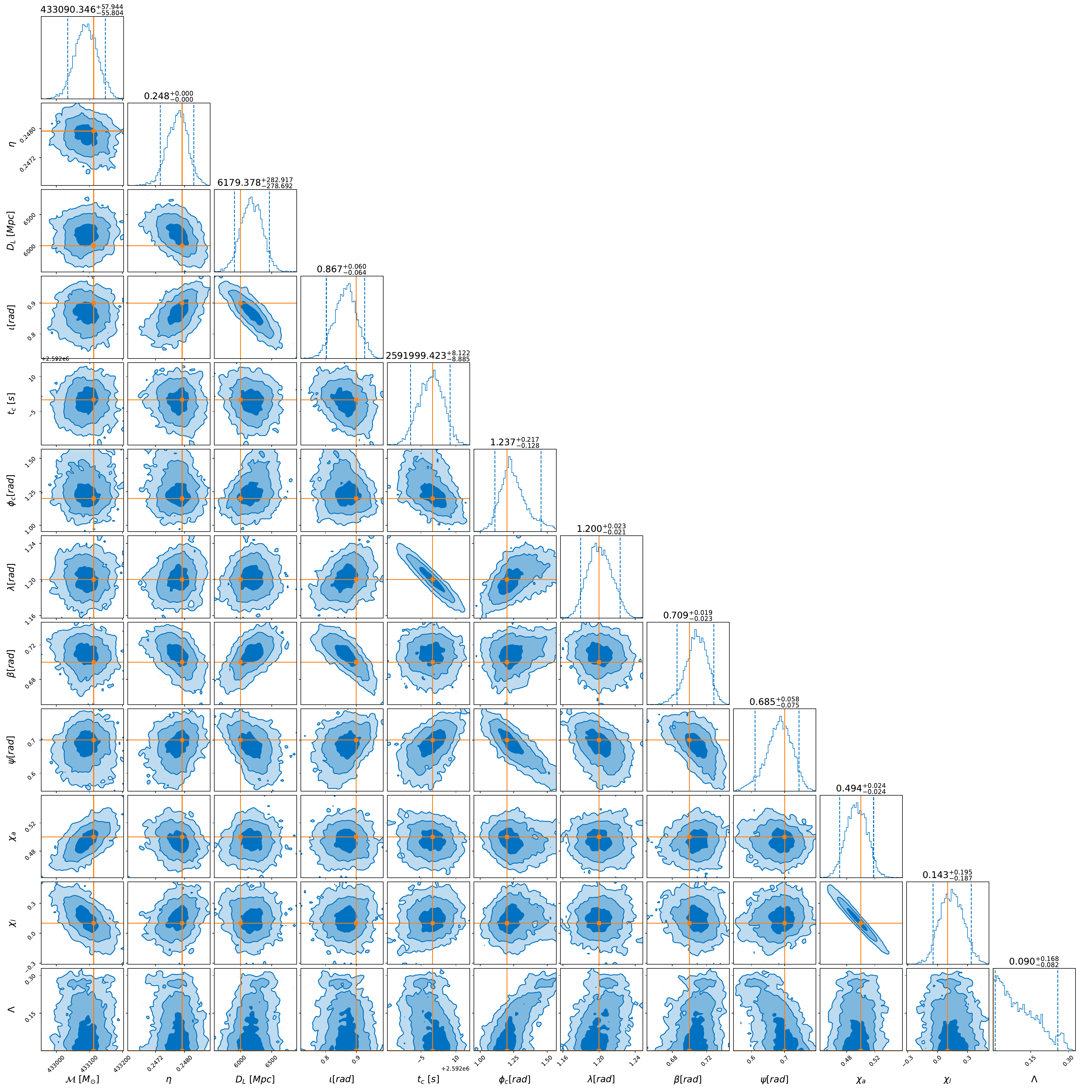}
	\caption{\label{fig:NC_1} Parameter posterior distribution for source I, with $1-2-3-\sigma$ contours and the orange line indicating the injected value. The two vertical dashed lines in each marginalized posterior distribution corresponds to is 90\% intervals.}
\end{figure*}

\begin{figure*}[t]
	\centering
	\includegraphics[width=\textwidth]{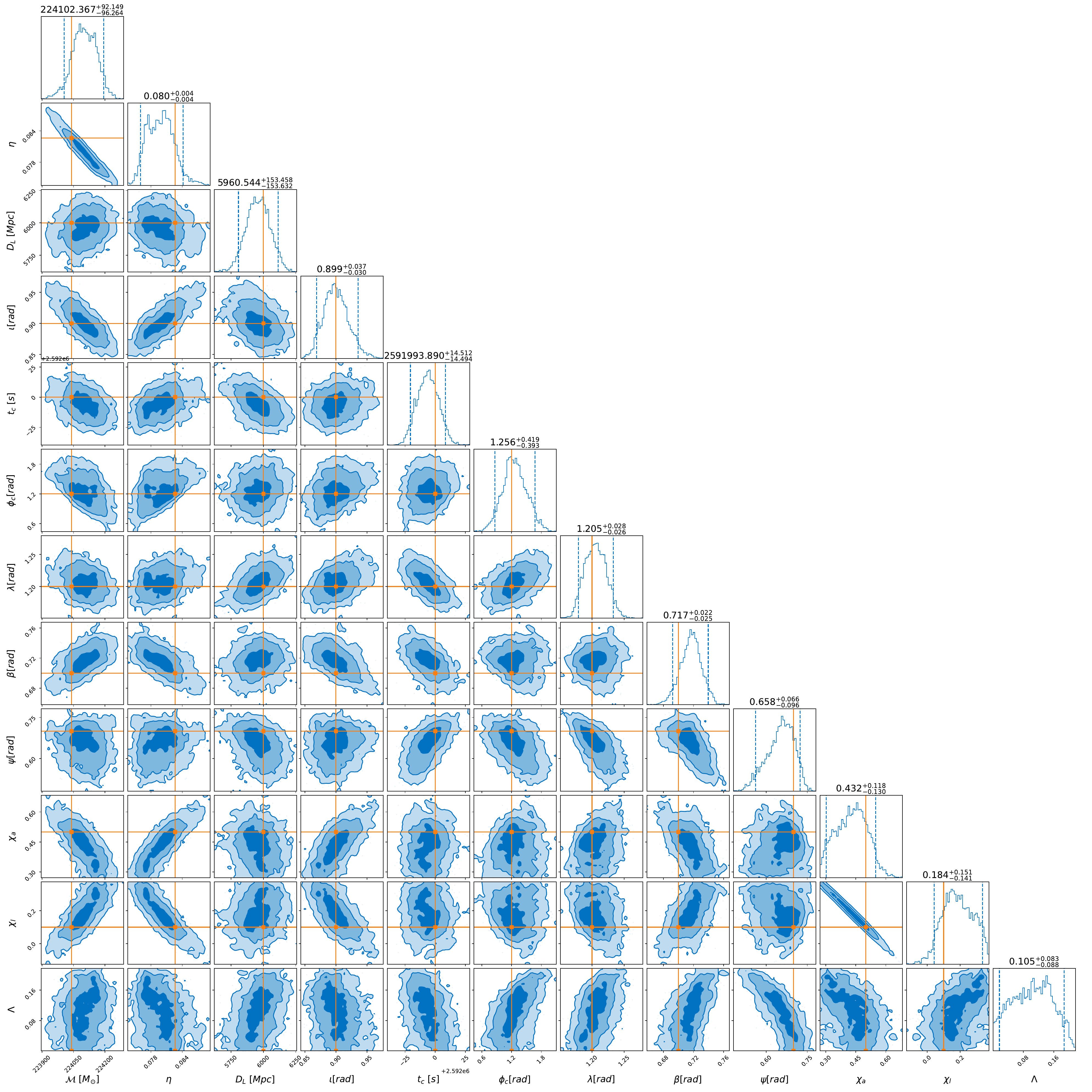}
	\caption{\label{fig:NC_2} Same as Fig.~\ref{fig:NC_1} but for the source II}
\end{figure*}

The detailed 12-dimensional posterior distributions for the two sources are displayed in Fig.~\ref{fig:NC_1} and Fig.~\ref{fig:NC_2}, respectively.
A thorough examination of these figures reveals a significant correlation between the spin parameters $\chi_a$
and $\chi_l$, it indicate that only the linear combinations of $\chi_1$ and $chi_2$ could not break the degeneracy. Nevertheless, it's worth noting that this reparameterization of spin parameters does enhance sampling efficiency and reduce computational time. Furthermore, the injected parameters value fall within the 90\% confidence intervals of the posterior distributions of the each parameters, and the parameter estimation accuracies are well align with Fisher's results.

\begin{figure}[b]
	\centering
	\includegraphics[width=0.5\textwidth]{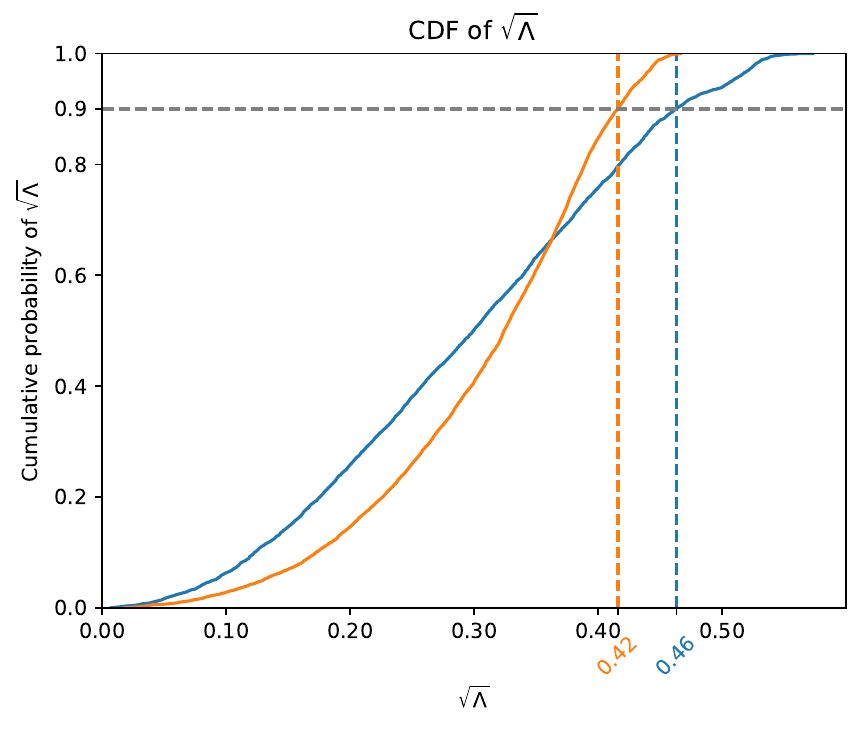}
	\caption{CDF of $\Lambda$. The blue line represents the result obtained from Source I, while the orange line represents the result from Source II. The two vertical lines indicate the locations where the accumulated probabilities reach 90\%.}\label{fig:CDF_lambda}
\end{figure}

The constraints of \ac{NC} are described by the cumulative probability distributions of the posteriors of $\sqrt{\Lambda}$. Fig.\ref{fig:CDF_lambda} provides a comprehensive overview of the results obtained from two simulated datasets. The blue line represents Source I, while the orange line represents Source II. Notably, the constraints at the 90\% confidence level for $\sqrt{\Lambda}$ are highlighted in the figure, corresponding to 0.46 and 0.42, respectively. For comparison, the results obtained using the Fisher information matrix are 0.36 and 0.29.

\begin{figure*}[t]
	\centering
	\includegraphics[width=\textwidth]{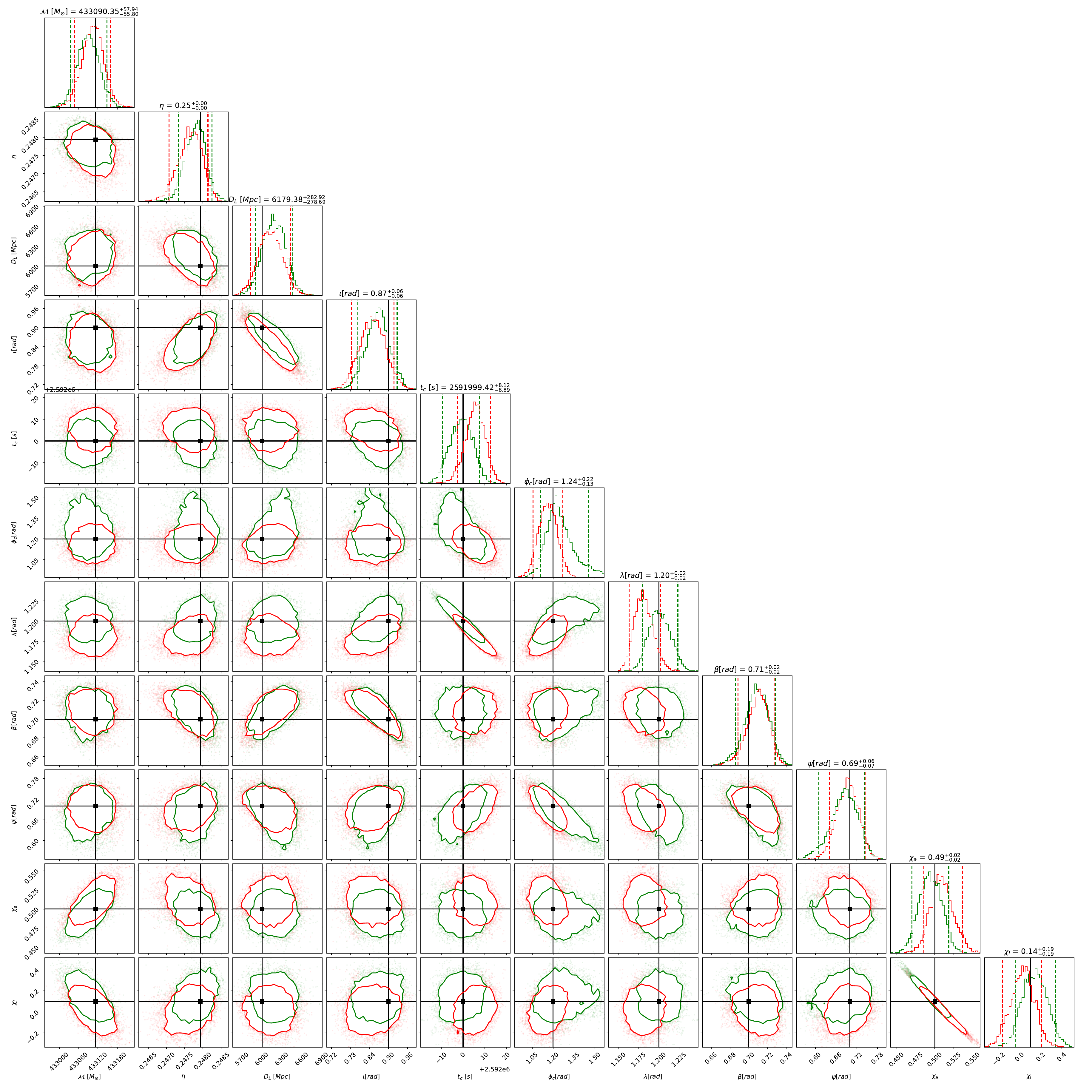}%
	\caption{\label{fig:150914_gr_nc} Comparison of the posterior distributions of the 11-dimensional source parameters under the GR model and the NC model, with the Source I. Red line is posterior distribution obtained by IMRPhenomXHM and green line is for NC model.}
\end{figure*}

We also examine the impact of non-GR parameter on the posterior distribution of source parameters. By comparing the posterior distributions of the 11-dimensional source parameters obtained using the GR model and the NC model, we found that the addition of parameter $\lambda$ does not significantly influence other GR parameters. As a case in point, Fig. \ref{fig:150914_gr_nc} presents the specific results for source I.

\section{Summary}\label{sec:summary}
In this paper, we have investigated the potential of TianQin in constraining NC using the inspiral signal from massive black hole binary. The findings suggest that TianQin has the capability to impose constraints on the temporal component of the non-commutative tensor that are at least one order of magnitude tighter than the Planck scale.

To address the future requirement for real data processing, we have developed a pipeline aiming at this topic for space-based \ac{GW} detector. To assess its performance, we have conducted Bayesian analysis on simulation data of two typical sources. The results show well alignment with Fisher information matrix, and a comparison of the posterior distribution of the \ac{GR} model with that of the space-time non-commutative model reveals that the additional parameters introduced by modified theories of gravity do not significantly impact the posterior distributions of the source parameters.

However, there are some limitations in this work. Firstly, the noise simulation for TianQin is based solely on the \ac{PSD}, and a more comprehensive and realistic noise model is required. Additionally, space-based \ac{GW} detectors face challenges such as multiple signal overlapping, further research need to investigate how to test \ac{GR} and constrain modified theories of gravity in this context.

\begin{acknowledgments}
	The authors thank Yi-Ming Hu and Jian-dong Zhang for useful discussions. This work is supported by the Guangdong
	Major Project of Basic and Applied Basic Research (Grant No. 2019B030302001)	and the National Science Foundation of China (Grant No. 12261131504).
\end{acknowledgments}


\bibliographystyle{apsrev4-1}
\bibliography{reference}

\end{document}